\documentclass{iopart}

\usepackage{iopams}

\usepackage{graphicx}

\begin{document}

\title[Fluctuation and melting of short DNA molecules]{Experimental
  and theoretical studies of sequence
  effects on the fluctuation and melting of short DNA molecules} 

\author{M. Peyrard$^1$, S. Cuesta L\'opez$^1$, D. Angelov$^2$}

\address{$^1$ Universit\'e de Lyon; 
Ecole Normale Sup\'erieure de Lyon, Laboratoire de Physique CNRS UMR 5672, 
46 all\'ee d'Italie,
  69364 Lyon Cedex 07, France
\ead{Michel.Peyrard@ens-lyon.fr}}

\address{$^2$ Universit\'e de Lyon; 
Ecole Normale Sup\'erieure de Lyon, Laboratoire de Biologie
  Mol\'eculaire de la Cellule CNRS UMR 5239, 
46 all\'ee d'Italie,
  69364 Lyon Cedex 07, France}

\begin{abstract}
Understanding the melting of short DNA sequences probes 
DNA at the scale of the genetic code and raises questions
which are very different from those posed by very long sequences,
which have been extensively studied. We investigate this problem by
combining experiments and theory. A new experimental method allows us
to make a mapping of the opening of the guanines along the sequence
as a function of temperature. 
The results indicate that non-local effects may be
important in DNA because an AT-rich region is able to influence the
opening of a base pair which is about 10 base pairs away. An earlier
mesoscopic model of DNA is modified to correctly describe the time
scales associated to the opening of individual base pairs well below
melting, and to properly take into account the sequence. Using this
model to analyze some characteristic sequences for which detailed
experimental data on the melting is available [Montrichok et
  al. 2003 Europhys. Lett. {\bf 62} 452], we show that we have to
introduce non-local effects of AT-rich regions to get acceptable
results. This brings a second indication that the influence of 
these highly fluctuating regions of DNA on their
neighborhood  can extend to some distance. 

\end{abstract}

\maketitle

\section{Introduction}

DNA is not the static double helix that structural pictures show. At
biological temperature its
thermal fluctuations are very large, leading to temporary openings of
the base pairs, and when temperature is increased up to $50$ or
$100\;^{\circ}$C, depending on the sequence, the two strands separate
in a transition often called ``DNA melting'' \cite{Wartell}. These
phenomena have been the object of many experimental and theoretical
studies in the last few decades but there are still open
questions. One of them is the understanding of the role of the
base-pair sequence. There are many aspects in this problem, depending on
the scale at which it is observed, i.e.\ the length of the DNA
molecule which is considered. Most of the earlier studies investigated
natural DNA molecules, with thousands or hundreds of thousands of base
pairs \cite{Wada}. Recording the melting curves of such DNA samples, for
instance through the variation of the UV absorbance of a solution,
analyzes the phenomena with a resolution of hundreds of base pairs or
more. Single molecule experiments observing the mechanical
denaturation of a long DNA segment \cite{Essevaz}, which is thermally
assisted, find a good correlation between the richness of the sequence
in GC base pairs, harder to break than AT pairs, and the force
necessary to open the molecule, but the resolution is again of the
order of 100 base pairs. 
Thus all these studies are not able to analyze sequence effects at the
scale of a few base pairs, which is the relevant scale for the genetic
code. 

On another hand it is now possible to perform a systematic study of
the fluctuations and melting of short DNA sequences (typically 20 to
60 base pairs), which can be
specifically designed to study a particular effect and
synthesized. This allows new experimental studies, which, in turn
raise interesting theoretical questions. This is the object of the
present paper.

Section \ref{sec:experiments} presents experimental results. Some of them
have been obtained by other authors but they are briefly reviewed here
because they provide interesting test results for the theoretical
analysis presented in Sec.~\ref{sec:theory}. Others are new and
demonstrate that the large fluctuations of AT-rich regions of DNA are
affecting the properties of the molecule several base pairs away. This
result must be taken into account to satisfactorily describe the
melting of some DNA sequences with a mesoscopic model, as discussed in
Sec.~\ref{sec:theory}. The study of some sequences which pose
a particular challenge to the theories of DNA melting
confirms the need to include the non-local effects of the AT-rich segments
of DNA to properly model the thermal denaturation. This gives
a second indication that 
the influence of highly fluctuating regions of DNA on their
neighborhood can extend to some distance.

\section{Experimental studies.}
\label{sec:experiments}

\subsection{Lifetime of the base pairs and melting curves}

The famous double-helix structure of DNA is made of two strands that
carry organic bases that are bound in pairs by hydrogen bonds, 
keeping the two strands together. There are  4 types of bases, called
A, T, G, C, but only  AT and GC pairs are part of the structure of DNA.
The AT pairs are bound by two
hydrogen bonds, while the stronger GC pairs are bound by three hydrogen bonds.
The covalent bonds that bind the atoms in the backbones and 
inside the bases are
very strong whereas the hydrogen bonds that connect the two bases in a pair
are much weaker. They can be broken by thermal fluctuations at
biological temperature, exposing the bases to the surrounding solvent
before the base pair closes again. These large fluctuations
which are called the ``breathing of DNA'' are well known by
biologists.
Early studies of these fluctuations relied on deuterium-proton
exchange. If DNA is dissolved into deuterated water, when the opening
of a base pair exposes to the solvent
the protons that form the hydrogen bonds, 
the so-called imino protons, those protons
can be exchanged with deuterium from the water molecules of the
solvent. The exchange rate can be accelerated by a catalyst. Then NMR
can be used to detect the deuterium atoms within the 
DNA molecule \cite{Leroy}. Kinetic  experiments
show that the lifetime of base pairs (time during which they stay
closed) is in the range of milliseconds at $35^{\circ}$C and 10 times
more at $0^{\circ}$C. 
At biological temperatures the experiments show that {\em
  single base pair opening events are the only mode of base pair
  disruption.} This indicates that the large amplitude conformational
change that is able to break a pair and expose a base to the solvent is
a {\em highly localized phenomenon}.  The lifetime of the open state,
estimated from kinetics data, is of a few tens of nanoseconds \cite{Leroy}.
At higher temperatures the
picture changes. Thermal 
fluctuations can break the base pairs in large regions of DNA, leading
to the so-called ``DNA bubbles'',  which grow when temperature is
raised, until 
the full separation of the strands. This {\em melting
  transition} of DNA is very sharp for homogeneous sequences. It can
be easily studied because the absorption of UV light at $219\;$nm
increases drastically when the base pairs are unstacked. The melting
transition shows up therefore as a sharp rise of the UV absorption of
the DNA solution, which only occurs within a few degrees for a short DNA
homopolymer.

For heteropolymers, such as DNA with a natural base sequence, the
melting profile is different because AT-rich regions tend to melt at lower
temperature than the GC-rich regions. This leads to melting curves
which are very sensitive to the details of the sequence, and not yet
fully understood although some empirical models can give rather
accurate predictions, at a resolution of a few hundreds of base pairs
for long sequences \cite{Wada}. For long molecules the thermal
separation of the two strands is probably well described by a zipper
mechanism. This is different for short DNA sequences of less than 100 base
pairs. Very short oligomers melt in a two-state process in which they
are either fully closed as double helices or fully separated in
individual strands. Longer molecules can exist in intermediate states
in which a part of the molecule is open and the other part is closed.
For instance they can have either open ends or open ``bubbles'' in the middle.
This raises a question to experimentalists. Measuring the UV
absorbance gives the fraction $f$ of open base pairs, but should a
fraction of say $0.5$  be interpreted as meaning that 50\% of the
molecules are fully closed and the other 50\% fully open, or does it
means that all the molecules are in an intermediate state where half
of their base pairs are open. This question can only be answered by
also determining the fraction $p$ of fully open molecules.

Using a clever trick, made possible by the availability of artificial
DNA sequences that can be tailored for a particular purpose, Montrichok et
al. \cite{Montrichok,ZENG} managed to do such a measurement. The idea is to use
sequences which are partly self-complementary. Each strand is such
that a set of bases at one end are complementary of the bases pairs at
the other end. As a result, when it is free, 
such a strand tends to fold into a hairpin,
the two ends forming a short DNA double helix closed by a loop made
by the central section of the strand, which does not include
self-complementary sequences. When a solution containing double
helices made of such strands 
is heated until a complete melting of all the molecules, 
and then cooled down, the strands
can either form hairpins or get together in pairs to reform the double
helix. But the formation of the hairpins from single strands is much
faster because two strands do not have to find each other by diffusing
in the solution. The experimental process used by Montricok et al. is the
following. They  slowly heat a solution of the DNA under study and
record its UV absorbance that gives $f$ versus temperature. 
At selected temperatures they
take an aliquot and cool it down quickly. All molecules which are fully
melted lead to the formation of hairpins while those which are only
partly open reclose as double helices. This allows them to determine
$p$ at these selected temperatures. 
This study \cite{Montrichok,ZENG} provides an
interesting set of experimental results on DNA melting because, from
the data, it is possible to extract the average length of the
denaturation bubbles as a function of temperature and the statistical
weight of bubble states. As discussed in Sec.~\ref{sec:theory}, owing
to peculiar properties of  some of the investigated sequences, these
experiments raise very challenging questions for the theoretical
analysis of DNA melting.

\subsection{Mapping of the opening along the sequence}

Proton-deuterium exchange measurements, and melting studies measuring
$f$ and $p$ provide a good set of experimental data of DNA
fluctuations and melting, but they lack an essential piece of
information, the spatial information that says precisely where the
fluctuations occur. This is why we performed a set of experiments,
using an original method, that can provide a mapping of the strength
of the fluctuations of the double helix as a function of the position
of along the sequence. This adds a new dimension to the melting
curves. 

\medskip
Usually extracting local information on DNA fluctuations
relies on a  dye or fluorophore probe inserted in a special
molecular construct, which reports on the properties of DNA
in its vicinity \cite{BRAUNS,BONNET2003}. This method has two major
drawbacks. First it provides information on a single point in the
sequence, the place where the probe is located. Second the probe is
probably not innocuous and it can perturb the dynamics of DNA in its
vicinity. {\em Our method asks DNA itself to report on its internal
  state.} It uses the propensity of the guanine bases G to be
ionized by biphotonic excitation from a strong UV laser pulse.
Two guanine modifications, oxazolone and
8-oxo-7,8-dihydro-2-oxoguanine (8-oxodG) have been identified as the major
one-electron oxidative DNA lesions. Their formation depends on the local
DNA conformation and charge-transfer efficiency
\cite{Douki,Angelov,Spassky97} which is affected by local
fluctuations. While oxazolone 
is the unique product resulting from one-electron oxidation of the
free $2'$-deoxyguanosine, 8-oxodG appears as soon as the nucleoside is
incorporated in a helical structure. Hence the measurement of the
relative yield of these photoproducts at each G site tells us whether
this G was in an helical structure (closed) or whether it was open
when the molecule was hit by the laser pulse. Therefore it gives a
snapshot of the opening of DNA at each G site. As the experiment is
not performed on a single molecule but in solution the results are
obtained on a statistical ensemble and they give the opening
probability at each G site at the temperature of the study.

\medskip
To measure the relative yield of the production of oxazolone and
8-oxodG we can rely on standard biological methods. Piperidine cleaves
DNA at the sites where it has oxazolone, i.e. where the guanine was
unstacked (open base pair), whereas Fpg protein cleaves it as sites
which have 8-oxodG. One end of the molecule is marked by a radioactive
marker, and, after cleavage by piperidine or Fpg, performed on
different aliquots at the same temperature, 
the lengths of the fragments that carry the marker
are measured by gel electrophoresis. The labelled
fragments of length $\ell$ are those which were cleaved at 
a guanine situated at distance $\ell$ from the marked end.
The ratio 
 $R_{Fpg}/R_{pip}$ of the number of radioactively labeled
fragments of length $\ell$ produced by Fpg cleavage or by piperidine
cleavage indicates the probability of closing for
this particular guanine. As the gel image contains the data for 
fragments of all lengths, a single experiment measures the opening
probability at all guanine sites.

\begin{figure}
  \includegraphics[width=12cm]{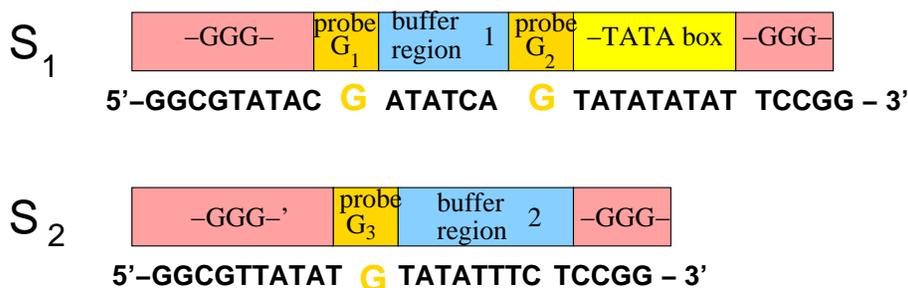}
  \caption{The two DNA sequences investigated by UV ionization of the guanines.}
  \label{fig:seqUV}
\end{figure}

\medskip
Since our goal was to study sequence effects we performed these
studies on two artificial sequences, specially designed to investigate
the role of an AT-rich region in DNA. Such AT-rich regions, called
TATA boxes, exist in the transcription-initiation regions of the genes
of various species. As the AT base pairs bound by only two hydrogen
bonds are weaker than the GC pairs, they are expected to exhibit large
fluctuations even at biological temperature because they are closer to their
melting temperature. Depending on its length a poly-AT DNA melts around
55$^{\circ}$C while a poly-GC melts around 80$^{\circ}$C. Figure
\ref{fig:seqUV} shows two sequences that we investigated. Sequence
$S_1$ contains a TATA box and two guanines, highlighted in the
sequence, that were monitored as probes of the fluctuations of the DNA
molecule. Probe $G_1$ is separated from the TATA box by a buffer
region which contains 7 bases, while
probe $G_2$ is adjacent to the TATA box, in the $5'$ direction of the DNA
strand. In sequence $S_2$ probe $G_3$ is surrounded by two small
AT-rich regions but the larger TATA box and the guanine $G_2$ have been
eliminated.
In both cases these short DNA molecules are terminated by
GC rich domains which act as clamps to prevent large fluctuations of
the free ends of the molecules, and hold the two strands together even
when we heat the sample up to 60$^{\circ}$C. In principle the guanines
in these terminal parts could also be monitored in our analysis, but the
spatial resolution near the end of the strands is not very good so
that it is hard to separate two adjacent guanines in this region.
This is why we have not studied them.

\bigskip

\begin{figure}
\begin{tabular}{cc}
\textbf{(a)} & \textbf{(b)} \\
\includegraphics[width=6cm]{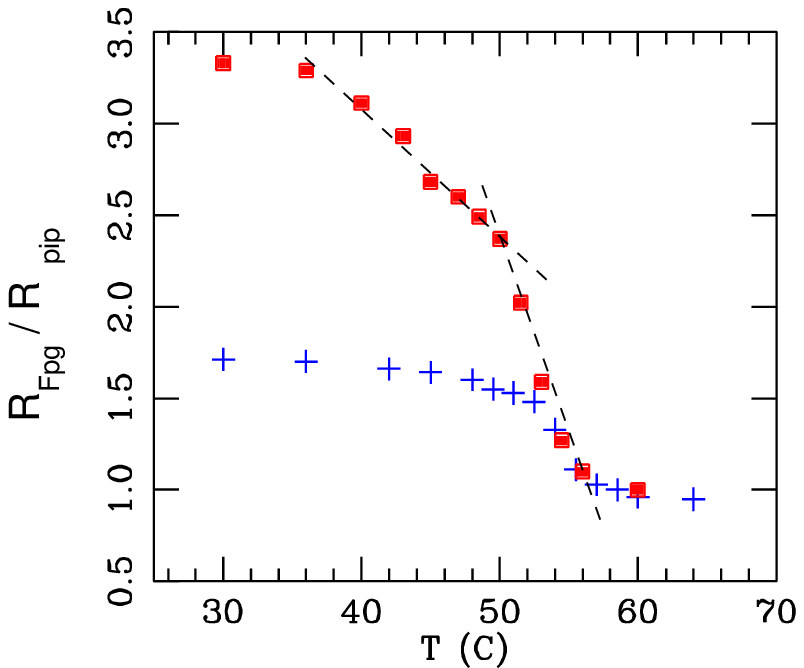} &
\includegraphics[width=6cm]{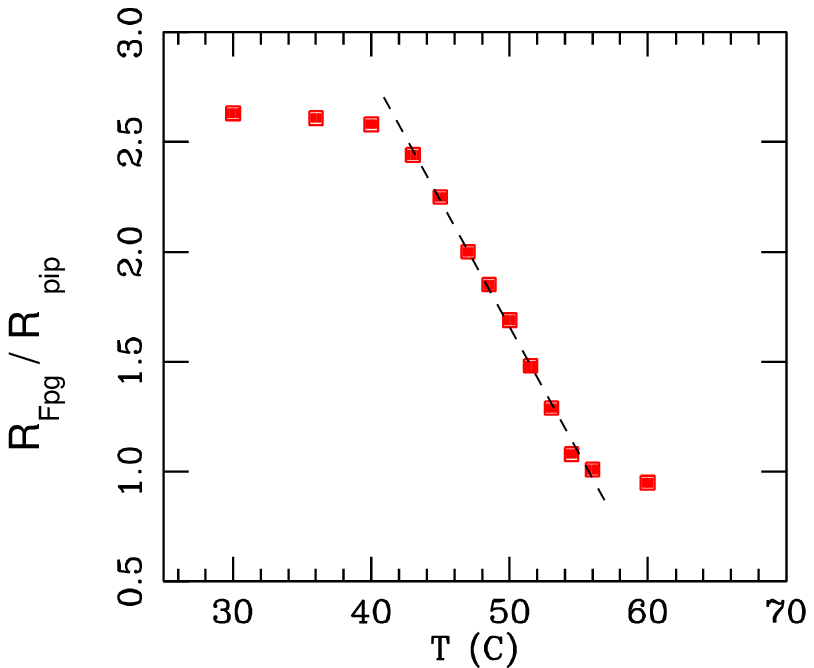} 
\end{tabular}
  \caption{Temperature dependence of the ratio $R_{Fpg}/R_{pip}$ for
    the guanine probe $G_1$ (squares) and $G_2$ (crosses) in sequence
    $S_1$ (a) and for the guanine probe $G_3$ in sequence $S_2$
    (b). The dotted lines are guides to the eye, pointing out
    the peculiarity of the curves. }
\label{fig:resuUV}
\end{figure}

\begin{figure}
\begin{center}
  \includegraphics[width=7cm]{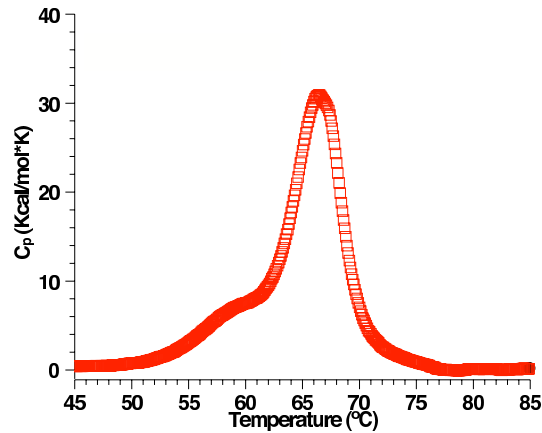}
\end{center}
  \caption{Excess molar specific heat of a DNA sample with sequence
    $S_1$ versus temperature. The large peak is associated to
    melting. Precursors effects appear as the shoulder below the
    transition temperature. In this experiment the solvent and 
    concentrations are
    different from those used in the UV ionization experiments,
     so that the transition temperature is shifted.}
  \label{fig:calo}
\end{figure}

Figure \ref{fig:resuUV} shows the variation versus temperature of
$R_{Fpg}/R_{pip}$ for the three guanine probes $G_1$, $G_2$, $G_3$.
The value of $R_{Fpg}/R_{pip}$ should not be considered as a
precise quantitative measure of the local closing probability 
because it is also
affected by the configuration of the DNA molecule near the probe,
which depends slightly on the sequence. Only the temperature
dependence of this ratio for a given probe can be analyzed
quantitatively. However the very low value of  $R_{Fpg}/R_{pip}$ for
probe $G_2$ next to the TATA box is a strong signal of very
large fluctuations occurring at the level of guanine $G_2$, even at low
temperature. When temperature is raised 
$R_{Fpg}/R_{pip}$ drops sharply around $55^{\circ}$C
for probe $G_2$, at the melting transition of the DNA molecule. The
variation of $R_{Fpg}/R_{pip}$ for probe $G_1$ in sequence $S_1$
(squares on Fig.~\ref{fig:resuUV}) is interesting because it evolves
in two stages. The first stage is a gradual linear decrease of the
closing probability of this guanine which starts at temperatures as
low as $36^{\circ}$C and extends to $50^{\circ}$. Above this
temperature, the second stage is a sharp decrease of the closing
probability of probe $G_1$ which is due to the melting transition, and
occurs at the same temperature as for probe $G_2$. Combining the data for
probes $G_2$ and $G_1$ suggests the following image of the fluctuation and
melting of a DNA sequence such as $S_1$: the TATA box exhibits
large fluctuations even at temperatures as low as $30^{\circ}$ which
strongly affect a base which is right next to it (probe $G_2$); as
temperature rises, these fluctuations extend farther away and start to
affect a base such as probe $G_1$ although is is separated from the
TATA box by a buffer region of 7 base pairs. This indicates that the
TATA box causes significant precursor effects below the
melting transition. This is confirmed by the calorimetric results
shown in Fig.~\ref{fig:calo}. Those precursor effects appear as a small
shoulder in the low-temperature side of the peak in specific heat
associated to the melting transition. The role of the TATA box in
these precursor effects is demonstrated by the counter example of
sequence $S_2$, which does not have a TATA box and shows a simple
one-stage melting as shown by the signal given by probe $G_3$ in
Fig.~\ref{fig:resuUV}-b. The melting transition is broader than for
sequence $S_1$ and occurs at a slightly lower temperature
because the molecule is shorter, but it does not show the change of
slope observed for probe $G_1$ with the TATA box.

\bigskip
These experiments indicate that a large AT-rich region such has the TATA
box may have a profound effect on the fluctuations and melting of a
DNA sequence. Its large fluctuations lead to precursor effects, called
``pre-melting'' which had already been observed in other sequences
\cite{ERFURTH,CHAN90,CHAN93,MOVILEANU2002}. Our study brings the
spatial information that was not available before and it turns out to
be important because it shows that {\it a guanine which is 8 base pairs away
from the TATA box is affected by the large fluctuations occurring at
the TATA box} even at temperatures well below the melting
temperature. These results also show that a proper understanding 
of the melting of
short DNA sequences is much more demanding than the analysis of the
melting curves of long DNA molecules. In those long molecules
melting curves register the opening of patches which may extend over
hundreds of base pairs, which smoothes out most of the details of the
sequence. On the contrary the melting of short DNA sequences is able to
show subtle sequence effects and our experimental results suggest that
{\em non local} effects may have to be included in the analysis. The
theoretical study of Sec.~\ref{sec:theory} confirms this view.

\section{Mesoscopic model of DNA melting.}
\label{sec:theory}

The thermal denaturation of DNA has attracted the attention of
theoreticians for more than four decades \cite{Zimm,Wartell} and the
problem is not over. Besides its biological interest this transition
poses a fundamental problem because DNA is essentially a
one-dimensional system, and it is generally assumed that the assertion
that phase transitions do not exist in one dimension is of general
validity. This question has been raised from the earlier studies
\cite{Poland} which describe DNA as a sequence of two-state systems,
the base pairs, which can be either closed or open,
i.e. Ising-like. This simplified description is unable to properly
describe the entropy of the loops formed by the strands in the open
regions and the model has to be completed by an evaluation of the
statistical weight of a loop. This weight, which determines
the entropy contribution of the open regions, depends on the size of the
loop and can be written as $A s^k/k^c$ for an open region of $k$ base
pairs where $A$, $s$ and $c$ are constants. 
This dependence on the loop size introduces an effective long range interaction
which makes a one-dimensional transition possible. The nature of
the transition depends sensitively on $c$, a value of $c > 2$ leading
to a first order transition, which 
is consistent with the sharp transition observed
experimentally. However the value of $c$ is very hard to calculate
because it should take into account the topology of the molecule. 
The DNA strands that form the loops cannot overlap, i.e. have to be
described by a self-avoiding walk. This is not enough to
give $c>2$ but  a recent study \cite{Mukamel}
showed that the avoided crossing between
the loops and the rest of the chain brings $c$ above 2, thus
justifying theoretically the sharp transition which is observed
experimentally.

\medskip
The Ising models of DNA melting are well suited for long DNA
chains. The theory itself has been developed with these cases in mind
because the crossings between the loops and between the loops and the
rest of the molecule are much more likely to occur when loops of
hundreds of bases are formed, while they are irrelevant when one
considers denaturated regions of a few base pairs. These models are
also very convenient for practical calculations, due to their
simplicity, although the long range effect associated to loop entropy
is still not trivial to treat efficiently \cite{FixmanFreire,Yeramian}. Some
improvements in the evaluation of the loop entropy weight
\cite{Blossey} and the development of efficient algorithms
\cite{FixmanFreire,Yeramian} turned this approach into a tool allowing
practical applications \cite{DINAMELT,Carlon}. However for peculiar
short sequences, this approach may lead to qualitatively wrong results
(Sec.~\ref{subsec:challenge}) and it has to rely on a very large
number of parameters \cite{SantaLucia} which are hard to evaluate
because they correspond to probabilities rather than interaction energies.

\subsection{Nonlinear model for DNA melting}
\label{sec:model}

\begin{figure}[h!]
  \centerline{
\includegraphics[width=6cm]{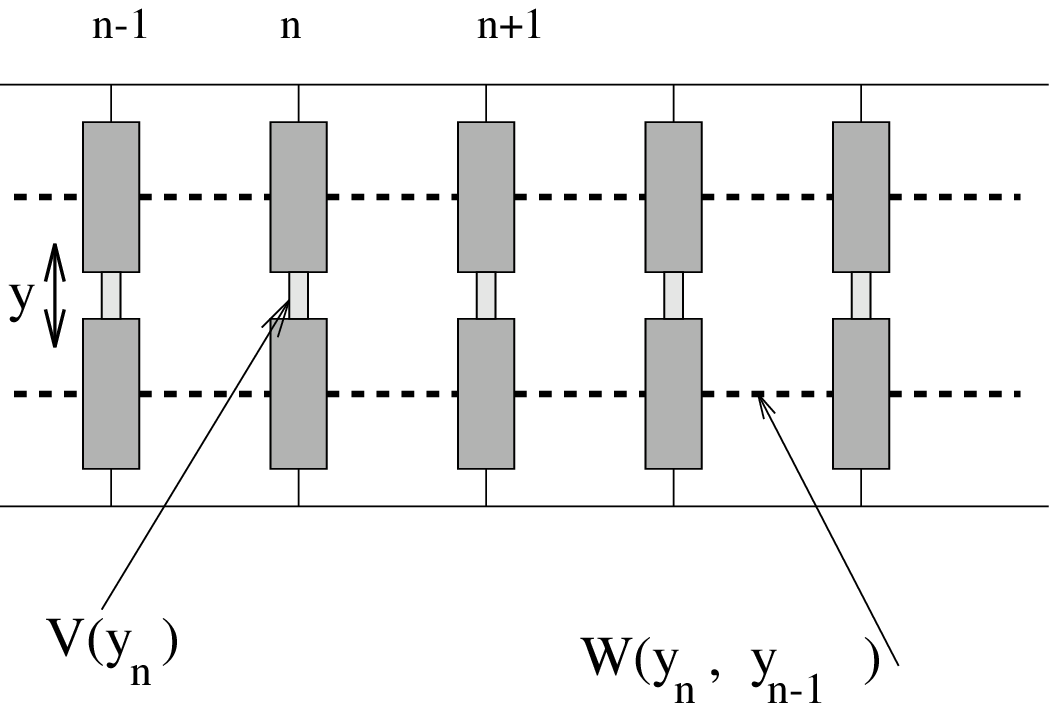}
\hspace{0.5cm}
\includegraphics[width=5.0cm]{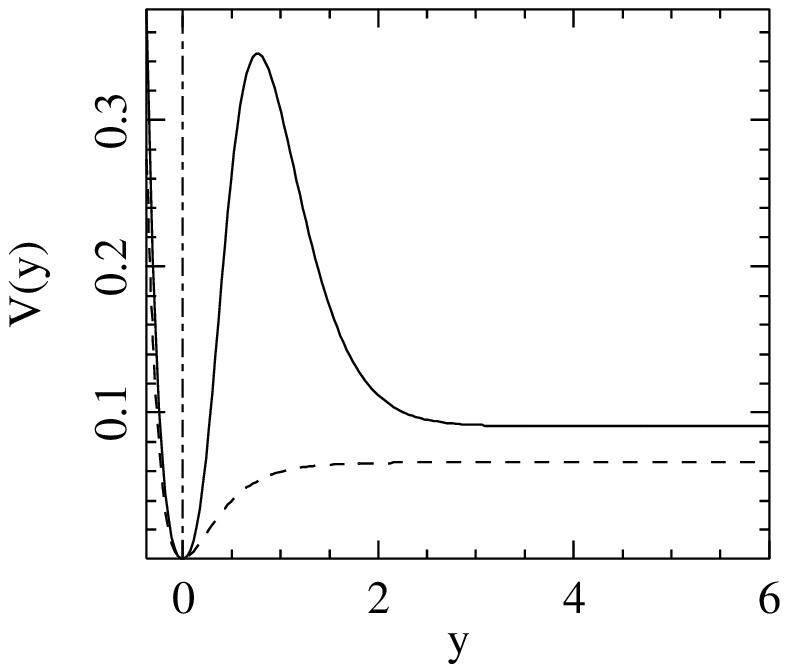}
}
\caption{The nonlinear DNA model described by
  Hamiltonian (\ref{eq:PBhamiltonian}). 
  The right panel shows the
  Morse potential (dotted line) and the potential defined by
  Eq.~\ref{eq:potmodmorse} (full line). The parameters of the two
  potentials are chosen such that the models using these potentials
  give the same denaturation temperature for a poly(A) DNA. }
\label{fig:PBmodel}
\end{figure}

This model \cite{MPreview,MP} can be viewed as a mesoscopic dynamic
model of
the DNA molecule, which ignores its helicoidal structure and
condensates all the atomic coordinates of a base pair into a single
number $y$ which describes the stretching the
bonds between the two bases. It
attempts to complete the Ising-model approach for DNA
because it not only considers the closed and open states of a base
pair but also all intermediate states, which allows a description of the
dynamics of the base pairs fluctuations. This is useful because it
brings additional information to validate the model and calibrate its
parameters. The model, schematized in Fig~\ref{fig:PBmodel}, is
defined by its Hamiltonian
\begin{equation}
  \label{eq:PBhamiltonian}
H = \sum_n \frac{p_n^2}{2m} + W(y_n,y_{n-1}) + V(y_n), \quad{\mbox{
    with}}\quad p_n = m \frac{dy_n}{dt} \; ,
\end{equation}
where $n$ is the index of a base pair and $m$ its reduced mass. 

The potentials $ W(y_n,y_{n-1})$ and $V(y_n)$ are the crucial elements
that define the model. In the original version of this model \cite{PB}
they were chosen as the simplest expressions leading to qualitatively
satisfactory functions, but this choice is not appropriate to
study the melting of specific DNA sequences, and they had to be
improved.

The potential $W(y_n, y_{n-1})$ describes the interaction between adjacent
bases along the DNA molecule. It has several physical origins:
\begin{itemize}
  \item the presence of the sugar-phosphate strand, which is rather rigid and
  connects the bases. Pulling a base out of the stack in a translational
  motion tends to pull the neighbors due to this link. One should notice
  however that we have not specified the three dimensional motion of the bases
  in this simple model. An increase of the base pair stretching could also be
  obtained by rotating the bases out of the stack, around an axis parallel to
  the helix axis and passing through the attachment point between a base and
  the sugar--phosphate strand. Such a motion would not couple the bases
  through the strands. The potential $W(y_n, y_{n-1})$
  is an effective potential which can be viewed as averaging over the
  different possibilities to displace the bases.
\item the direct interaction between the base pair plateaus, which is due to
  an overlap of the $\pi$-electron orbitals of the organic rings that make up
  the bases.
\end{itemize}
In a first step \cite{PB} a harmonic interaction $W(y_n, y_{n-1})$ was used, but
this is a crude approximation since experiments show that a base pair
can open independently of its neighbors. These large relative
displacements rule out a low order expansion of the potential
and the nonlinearity of the
stacking interaction should not be ignored. The shape
of the potential is also constrained because the
model must reproduce the sharpness of the melting
transition of DNA. This sharpness is associated with an entropic
effect. Opening the base pairs has an energy cost
which is the energy
required to break the bonds in the pairs, 
but there is an entropy gain because,
once they are opened the bases are freer to fluctuate. It is this effect
which allows a transition in a one-dimensional system like DNA
\cite{TDPtransition}. The model can lead to realistic melting
curves if the potential $W(y_n, y_{n-1})$ takes into account the extra
freedom of the broken base pairs. This can be described \cite{DPB}
by choosing
\begin{equation}
 \label{eq:nlstacking}
W(y_n, y_{n-1}) = \frac{1}{2} K \left( 1 + \rho e^{- \delta (y_n + y_{n-1}) }
\right) (y_n - y_{n-1})^2 \;.
\end{equation}
This expression can be viewed as a harmonic interaction with a
variable coupling constant.
As soon as either one of the two interacting base pairs is open 
(not necessarily both simultaneously) the effective coupling constant
drops from $K' \approx K(1 + \rho)$ down to $K' \approx K$. The smaller
coupling leads to an entropy increase, which promotes the transition
by reducing the free energy of the open state.

\bigskip
The potential $V(y)$ describes the interaction between the two bases in a
pair. The first nonlinear model of DNA melting \cite{PB,MP} 
used a Morse potential
\begin{equation}
  \label{eq:MorsePot}
V(y) = D \left( e^{-\alpha y} - 1 \right)^2 \; ,
\end{equation}
where $D$ is the dissociation energy of the pair and $\alpha$ a parameter,
homogeneous to the inverse of a length, which sets the spatial scale of the
potential. This expression had been chosen because it is a standard expression
for chemical bonds and, moreover, it has the appropriate qualitative
  shape to describe the potential energy of the two bound bases:
 (i) it includes a strong repulsive part for $y<0$, corresponding to the
  steric hindrance,
 (ii) it has a minimum at the equilibrium position $y=0$,
 (iii) it becomes flat for large $y$, giving a force between the bases that
 tends to vanish, as expected when the bases are very far apart; this feature
 allows a complete dissociation of the base pair.
This potential, which can satisfactorily describe 
the equilibrium properties of DNA is however qualitatively wrong for the
dynamics. With the Morse potential 
both the lifetime of a base pair, i.e. the time
during which it stays closed between two opening fluctuations, and the
average time during which it stays open after such a fluctuation, are
found to be several orders of magnitude smaller than their experimental
values inferred from proton--deuterium exchange experiments
\cite{Leroy}. Entropic effects have to be taken into account in this
local potential $V(y_n)$ as well as in the coupling potential $W$ because,
irrespectively of the stacking interaction, when a base flips out of
the DNA stack, it gains new degrees of freedom such as a possible
rotation of
the plane of the base plateau, that were constrained in the helical
structure. But, in a mesoscopic model such as the one that we
consider, the ``potentials'' are actually potentials of mean force,
which take into account all the other degrees of freedom that are
ignored in the model, in a statistical way. The entropy gain that
follows the flipping of a single base out of the stack contributes to lower
the effective potential, whereas, in order to reclose the base has to
overcome a very large entropic barrier. For a correct dynamics of the
open states this entropic barrier has to be included in the potential
$V(y)$. 
The existence of this barrier has also been observed in free energy
calculations deduced from all-atom molecular dynamics simulations of
DNA \cite{Lavery}. Besides the entropic effect, a barrier for closing
might have a pure enthalpic contribution \cite{Drukker,Weber} because
the open bases tend to form hydrogen bonds with the solvent, which
have to be broken before closing.
We have chosen the expression
\begin{equation}
    \label{eq:potmodmorse}
V(y) = D \left( e^{-\alpha y} - 1 \right)^2  
+ \Theta(y) \frac{b y^3}{\cosh^2[ c(\alpha y - d \; \ln 2)]}\; ,
\end{equation}
where $\Theta(y)$ is the Heaviside step function which ensures that the
term added to the Morse potential
only plays a role for $y>0$. This expression has been
chosen because it has the correct qualitative shape to generate the
entropic barrier, and, thanks to the factor $y^3$ the frequency at the
bottom of the potential is not affected by the additional
contribution. The parameters $b$ which determines the amplitude of the
barrier, $c$ its width, and $d$ its position in units of the value of $y$
at the inflexion point of the Morse potential, are constants for a given
type of base pair. Figure \ref{fig:PBmodel} compares the potential
(\ref{eq:potmodmorse}) and the Morse potential that give the same
denaturation temperature for a DNA homopolymer and a given stacking
potential $W$. For the parameters that we have chosen the entropy
barrier looks very high, but, one must keep in mind that 
an open base is also pulled back toward its closed state by the stacking
interaction with its closed neighbors, so that the barrier for
reclosing in the total effective potential
experienced by an open base is actually much smaller. The calculation
of the average lifetime of an open base can be used to determine the
value of the potential parameter $b$ which controls the barrier
height. 

\begin{figure}
  \centerline{\includegraphics[height=12cm,angle=270]{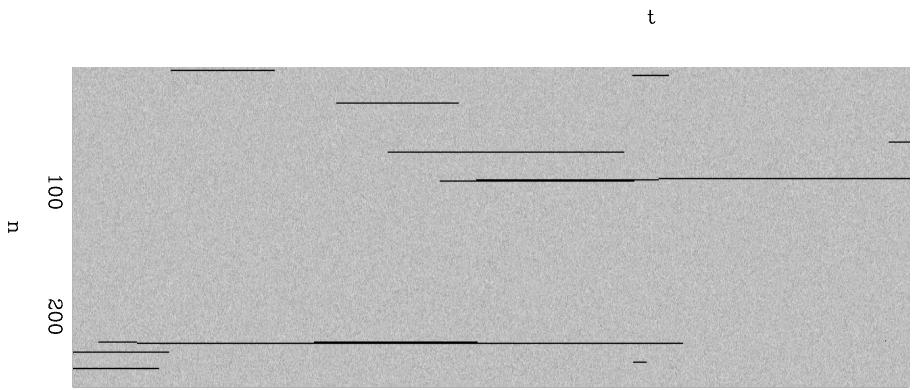}}
  \caption{Time dependence of the base pair opening in a  poly(A) DNA,
    according to a molecular dynamics simulation of the
    model. The parameters are those defined
    in Table~\ref{table:param} for AA,
$m = 300\;$ atomic mass units, and the temperature is $T=300\;$K.
The horizontal axis corresponds to time and extends over
$0.4\;\mu$s. The vertical axis extends along the sequence which has
256 base pairs in this calculation. The stretching of each base pair
is indicated by a
grey scale, lighter grey corresponding to closed and back indicates a
base pair stretching $y > 1.5\;$AA, i.e. an open site. Therefore the
horizontal black lines correspond to open sites, that stay open for a
long time. } 
  \label{fig:grey}
\end{figure}

Figure~\ref{fig:grey} shows an example of the dynamics of the model of
a poly(A) DNA at $300\;$K. Most of the base pairs stay closed
during the whole simulation, but a few of them open, and stay open for
a long time, before closing again. These events concern a single pair,
as observed experimentally \cite{Leroy}. Longer calculation find an
average opening time of $40\;$ns, in the experimental range. 
The lifetime of a closed pair is
difficult to evaluate quantitatively because it is longer that the time
interval that we simulate. Figure ~\ref{fig:grey} shows that, if we
observe the system for $0.4\;\mu$s,  7\% of the bases open,
leading to a lifetime of a closed pair of the order of $6\;\mu$s. However
this is only an estimate which is very sensitive to the
model parameters, and of course to temperature. Although this value is
several orders of magnitude higher that the value given by a Morse
potential, it is still much smaller than the experimental estimate of
a few ms. The oversimplified model that we use can certainly partly
explain this discrepancy, but it may also be attributed to differences
between the quantities which are actually observed. Experiments detect
proton-deuterium exchange and may miss short openings, which in our
numerical observation are counted and reduce the estimated time during
which a base pair stays closed.

\subsection{Computing DNA melting curves}
\label{subsec:computing}

Once the model is selected, the next step is to use it to compute the
melting curves of short DNA sequences. The simplest approach would be
to run molecular dynamics simulations of the equations of motions that
derive from Hamiltonian (\ref{eq:PBhamiltonian}), in the presence of a
thermal bath to control temperature. Recording the fraction of
denaturated base pairs, defined as the base pairs for which the
average value of $y$
exceeds a threshold $\xi$ larger than the stretching of a base pair
at the maximum of the entropic barrier of $V(y)$ (we henceforth
use $\xi = 2\;$\AA), one could expect to compute the melting
curve. This is however a bad choice for two reasons. First, in order
to obtain good statistics very long simulations, or a very large
number of realizations, would be required. But there is a more
fundamental problem. The denaturation transition does exist for an
infinite system, but, for a finite DNA molecule the model leads to a
full denaturation at any temperature. This time can be very long for
long DNA molecules, but simulation results are nevertheless
meaningless. The reason behind this behavior is that the model
describes an infinitely dilute DNA solution. A complete picture should
include the hybridization of two separated strands coming together in
solution and combining to form the double helix. This equilibrium
between the dissociated strands and the double helix can, in
principle, be described by an external contribution in the partition
function of the system \cite{Wartell}, but it is difficult to evaluate
quantitatively to a reasonable accuracy. To get unambiguous results we
{\em must} restrict our attention to a reduced statistical ensemble,
the dsDNA ensemble \cite{vanErp}, which is defined as the ensemble of
the molecules which have {\em at least} one base pair which is formed,
i.e.\ with $y < \xi$. In this ensemble the fraction of open base pairs
$\sigma(T)$ is well defined and can be computed exactly with the
methods of statistical physics, which do not suffer from the sampling
limitations of molecular dynamics calculations. It is interesting
that, thanks to the experimental method of Montrichok et
al. \cite{Montrichok,ZENG} the denaturation curve in the dsDNA
ensemble can be measured experimentally, from the fraction $f(T)$ of open
base pairs, irrespectively of the state of the molecule, recorded for
instance by UV absorption spectroscopy, and the fraction $p(T)$ of
molecules which are fully melted because $f(T)$ is given by
\begin{equation}
  f(T) = [ 1 - p(T)] \sigma(T) + p(T) \; .
\end{equation}

\bigskip

An exact theoretical calculation of $\sigma(T)$ is possible because
the model is one-dimensional and only includes nearest neighbor
coupling, so that the calculation of its partition function can be
done by direct integration \cite{vanErp}. For a DNA molecule which
comprises $N$ base pairs it reduces to a sequence of one-dimensional
integrals over the variables $y_1 \ldots y_N$ which can be performed
by a simple iterative scheme. Let us give a sketch of the process (for
a complete discussion see \cite{vanErp}). Let
\begin{equation}
  Z_I = \int \Pi dy^N e^{-\beta U(y^N)}
\end{equation}
be the configuration integral performed over an infinite domain for
each $y_i$ (in practice the calculation has to be done over a finite
domain, with an appropriate cutoff \cite{vanErp}), where $y^N$ denotes
the ensemble of all the variables $y_1 \ldots y_N$ and $ U(y^N)$ is
the potential energy of the model, sum of the $W$ and $V$
contributions. Similarly we can define
\begin{equation}
  Z_{II} = \int_{y^N > \xi} \Pi dy^N e^{-\beta U(y^N)}
\end{equation}
to be the same integral with the condition that {\em all} the $y_i$
are simultaneously larger than $\xi$. $Z_{II}$ gives the statistical
weight of the fully open states of the model so that the configuration
integral of the dsDNA ensemble is
\begin{equation}
  Z = Z_I -  Z_{II} \; ,
\end{equation}
which is well defined and does not depend on the upper cutoff. To get
for instance the statistical weight of the states for which base pair
$j$ is closed we have to compute
\begin{equation}
  Z (j \;{\rm closed})  = \int_{y_j < \xi} \Pi dy^N e^{-\beta U(y^N)}
\end{equation}
where the variable $y_j$ is constrained to be smaller than $\xi$, the
others being unconstrained. This gives the probability that base pair
$j$ is closed in the dsDNA ensemble as
\begin{equation}
  P_{dsDNA}(j \; {\rm closed}) = \frac{Z (j \; {\rm closed})}{ Z_I -
    Z_{II}} \; ,
\end{equation}
from which the dsDNA melting curve can be obtained as well as a
three-dimensional plot showing $ P_{dsDNA}(j \; {\rm closed},T)$ versus
$T$ for all $j$. This figure is the generalization of the melting curve by
the addition of the spatial dimension. It is the theoretical
equivalent of the experimental result that we can obtain, for guanine
sites only, by the UV-ionization method.

\subsection{The challenge posed by some sequences}
\label{subsec:challenge}

The last step to analyze theoretically the melting curves of short DNA
molecules is to introduce the sequence in the model. This is naturally
done by considering sequence dependent parameters. The first studies
using the PBD model (model with a Morse potential $V(y)$) to describe
experimental denaturation curves of short DNA molecules \cite{CAMPA}
assumed that the sequence is only contained in the on-site potential
$V(y)$ and that the stacking interaction $W$ is sequence
independent. Indeed, because the GC base pairs are linked by three
hydrogen bonds while the AT pairs are linked by only two, it seemed
reasonable to assume $D_{GC} = \frac{3}{2} D_{AT}$. However one must
keep in mind that $V(y)$ is an {\em effective} potential which
includes many contributions. Even if one ignores the entropic effects
that we introduced in Eq.~(\ref{eq:potmodmorse}) there are
features of DNA  other than the hydrogen bonds
which contribute to determine the potential that holds
the two strands of DNA together. The most
important is the electrostatic repulsion between the highly charged
phosphate groups of the strands, which is screen by counterions in
solution. Without these ions the double helix would not be stable at
biological temperature. The electrostatic repulsion explains why the
binding energies $D_{GC}$ and $D_{AT}$ determined in the first fits of
DNA melting curves with the PBD model were much smaller than typical
hydrogen bond energies. As the phosphate repulsion, which contributes
significantly to $V(y)$, is almost independent of the sequence, the
variation of $V(y)$ with the type of base pairs can be expected to be
much smaller than a $\frac{3}{2}$ ratio. Some analysis of
thermodynamic experiments on different DNA sequences \cite{GOTOH} do not
even include the potential $V(y)$. All these data indicate that,
although $V(y)$ may vary slightly from GC to AT base pairs, this
potential alone cannot describe all the effects of the sequence, which
must also be included in the stacking interactions $W$. Another fact
that supports this idea is the large range in which the stacking
energies vary. If we consider for instance the values given by quantum
chemistry calculation \cite{Saenger} there is a ratio of about 6
between the smallest and the largest value. Such variations cannot
be ignored.

\begin{figure}[h!]
\begin{center}
\includegraphics[width=5cm]{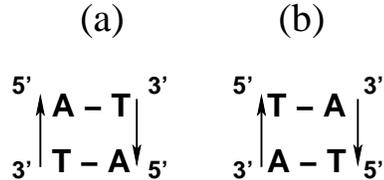}
\end{center}
  \caption{An example of base pair stacking in DNA showing why the
    orientation of the strands can lead to different stacking energies
  although they involve the same couple of base pairs. In the atomic
  structure of DNA the relative positions of the base pairs are
  different in cases (a) and (b), resulting in different interaction energies.}
 \label{fig:stackAT}
\end{figure}

\medskip
Including the sequence in stacking interactions $W$ is necessarily
more difficult than for the intra-pair potential $V$ because while $V$ can
only have two forms, for the AT and GC pairs, there are many more
possibilities for the stacking interactions. Two types of base
pairs lead to 4 possibilities, but this is an oversimplified view
because the interaction potential of AT over AT is different from that
of AT over TA. Even though the model does not explicitely take into
account the two strands because it is restricted to one degree of
freedom for a base pair, it is nevertheless possible to take into
account the sequence and distinguish between the two cases listed
above. The situation is even more complex because a DNA strand is
oriented. The geometry of the sugar phosphate backbone is not
invariant by reversal, and its orientation is conventionally defined
with a standard labeling of the carbon atoms on the sugar ring from
$5'$C to $3'$C. In the double helix the two strands are oriented in
opposite directions. Figure \ref{fig:stackAT} shows an example
involving two AT base pairs which are stacked in two different ways
along the sequence. Although these two cases involve the same couple
of base pairs, the relative position of the base pairs in the atomic
structure of DNA is not the same in the two cases so that their
interaction energies are significantly different, as measured for
instance by thermodynamic studies \cite{GOTOH}.
Taking into account the effect of the orientation of the strands gives
up to 16 possibilities for the stacking potential $W$. In the following
we label each of them by the initials of the two bases along one
strand, starting from the $5'$ end. Thus, for the two examples shown
in Fig.~\ref{fig:stackAT}, case (a) is labeled A--T and case (b) is
labeled T--A.

Considering all these possibilities immediately introduces a large
number of parameters in the model because the expresion of
the potential $W$ depends on
three parameters, the strength of the interaction $K$ (having the
dimension of an
energy divided by the square of a length), $\rho$ (dimensionless)
which measures the magnitude of the variation of the stacking when
base pairs open because the
effective constant drops from $K(1 + \rho)$ to $K$ when at least one one of the
interacting base pairs opens, and $\delta$ (dimension of the inverse
of a length) which determines how large the opening of a base pair
should be for this variation to play a role. To reduce the
number of parameters, we decided to select the same value of $\rho$
and $\delta$ for all the stacking interactions. We set $\rho = 25$,
which gives a large decrease of the stacking when either of the
interacting base pairs opens. This is a necessary condition to have a
sharp denaturation transition in DNA \cite{TDPtransition}, 
in agreement with experiments, and a large value of $\rho$ is also
necessary to match neutron scattering experiments that probe the
length of the closed regions of DNA versus temperature \cite{neutrons}.
The value $\delta = 0.8$ was chosen because it
leads to a drop the factor $\exp(-\delta y)$, to
$\frac{1}{5}$ of its original value for a
stretching $y \approx 2\;$\AA\ of a base pair, corresponding to overcoming
the barrier in the intra-pair potential $V$.

Consequently the dependence of the sequence in the interaction is
entirely included in the variation of $K$. We selected a parameter set
based on the results of theoretical calculations of stacking energies
\cite{Saenger} and evaluations of the stabilities of DNA doublets
deduced from thermodynamic measurements. Table \ref{table:param} list
all the parameters that we selected for the model.

\begin{table}
\begin{center}
\begin{tabular}{|l|ccccc|}
\hline\hline
Potential $V$ & $D$ & $\alpha$ & $b$ & $c$ & $d$ \\
\hline
AT base pair & $0.09075\;$eV & $3.0\;$\AA$^{-1}$ & $4.00\;$eV &
$0.74\;$\AA$^{-1}$ 
& $0.20\;$ \\
GC base pair & $0.09900\;$eV & $3.4\;$\AA$^{-1}$ & $6.00\;$eV &
$0.74\;$\AA$^{-1}$  
&$ 0.20\;$ \\
\hline\hline
\end{tabular}

\bigskip
\begin{tabular}{|l|cccccc|}
\hline\hline
Potential $W$ & \multicolumn{3}{c}{$\rho = 25$} & \multicolumn{3}{c|}{$\delta
  = 0.8\;$\AA$^{-1}$} \\
\hline\hline
Dimer &  A--T & A--A &  T--T & G--T & A--C &  T--A \\
$K\;$(eV \AA$^{-3}$) & 0.00176 & 0.00418 & 0.00418 & 0.00480 & 0.00462&
0.00506 \\
\hline\hline
Dimer &  G--A & T--C &  C--C & G--G & G--C &  C--T \\
$K\;$(eV \AA$^{-3}$) & 0.00546 & 0.00546 &  0.00810 &  0.00810 &
0.00865 & 0.00865 \\
\hline\hline
Dimer & A--G &  C--A &  T--G &  C--G & & \\ 
$K\;$(eV \AA$^{-3}$) & 0.00865 & 0.01140 &  0.01140 & 0.01690 & & \\
\hline\hline
\end{tabular}
\end{center}
\caption{Potential parameters used in the model. }
\label{table:param}
\end{table}

The intra-pair potential is inspired from the results fitted for the
PBD model \cite{CAMPA}, but, as explained above the relative
difference between AT and GC pairs has been reduced. The average depth of the
potential $D$ has been slightly increased to get better results for
the denaturation curves that we tested but also because our analysis
of the fluctuations of DNA hairpins \cite{Errami} suggested that $D$
was underestimated. The entropic barrier was added with a magnitude
$b$ chosen to give
correct lifetimes of the open and closed states (Sec.~\ref{sec:model})
and a width and position ($c$ and $d$) estimated from the actual
geometry of DNA.

Once $V$ has been selected the values of the parameters $K$ for the 16
possible dimers can be obtained from the melting temperatures of DNA
made of these dimers \cite{GOTOH}. Using the transfer integral method
we can calculate the melting temperature for a homopolymer DNA
\cite{DPB2}. For each dimer we can determine a value of $K$ by
considering a ficticious homopolymer with the stacking interaction
corresponding to this dimer. Strictly speaking this is only valid for
dimers made of twice the same base, but this approach gives
nevertheless values of $K$ which lead to melting temperatures in
rather good agreement with the observations for sequences that are made
of this dimer repeated many times. These values of $K$, computed for
infinitely long molecules have to be corrected for short
segments. This can be done by direct integration of the partition
function as explained in Seq.~\ref{subsec:computing}.

\bigskip
Of course the parameters that we selected in this process should only
be considered as a good starting point for further refinements. From
the fit of many experimental melting curves, completed by the local
analysis of the fluctuations obtained from the UV-ionization
experiments, one could expect to design an optimal parameter set for
the model, in the spirit of what has been done for Ising models of
DNA. This would be a full research program, which is not completed
yet, but, before it can be initiated, there are other questions to be
solved. They are related to the non-local effect of the large
fluctuations in some DNA regions that have been revealed by the
experiments described in Sec.~\ref{sec:experiments}. They show up in
the investigations of some sequences which pose special difficulties. 
These challenging sequences are particularly
interesting to investigate because they give us precious
informations on the fluctuations of DNA.

One example of such sequences is provided by the sequence $L48AS$
studied in the work of Montrichok et al. \cite{Montrichok,ZENG}, for
which experimental data on the melting curve are available, but also
the measurement of the fraction of fully denaturated molecules, which
allowed the authors to compute what we call the dsDNA melting curve
$\sigma(T)$. 

\begin{figure}[h!]
  \includegraphics[width=\textwidth]{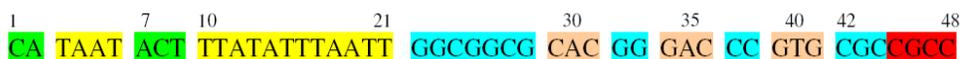}
  \caption{The sequence L48AS investigated by Montrichock et
    al. \cite{Montrichok,ZENG}. Only one strand is shown, the $5'$ end
    being on the left.}
\label{fig:seq48}
\end{figure}

\begin{figure}
  \begin{center}
    \begin{tabular}{cc}
\textbf{(a)} & \textbf{(b)} \\
\includegraphics[height=5cm]{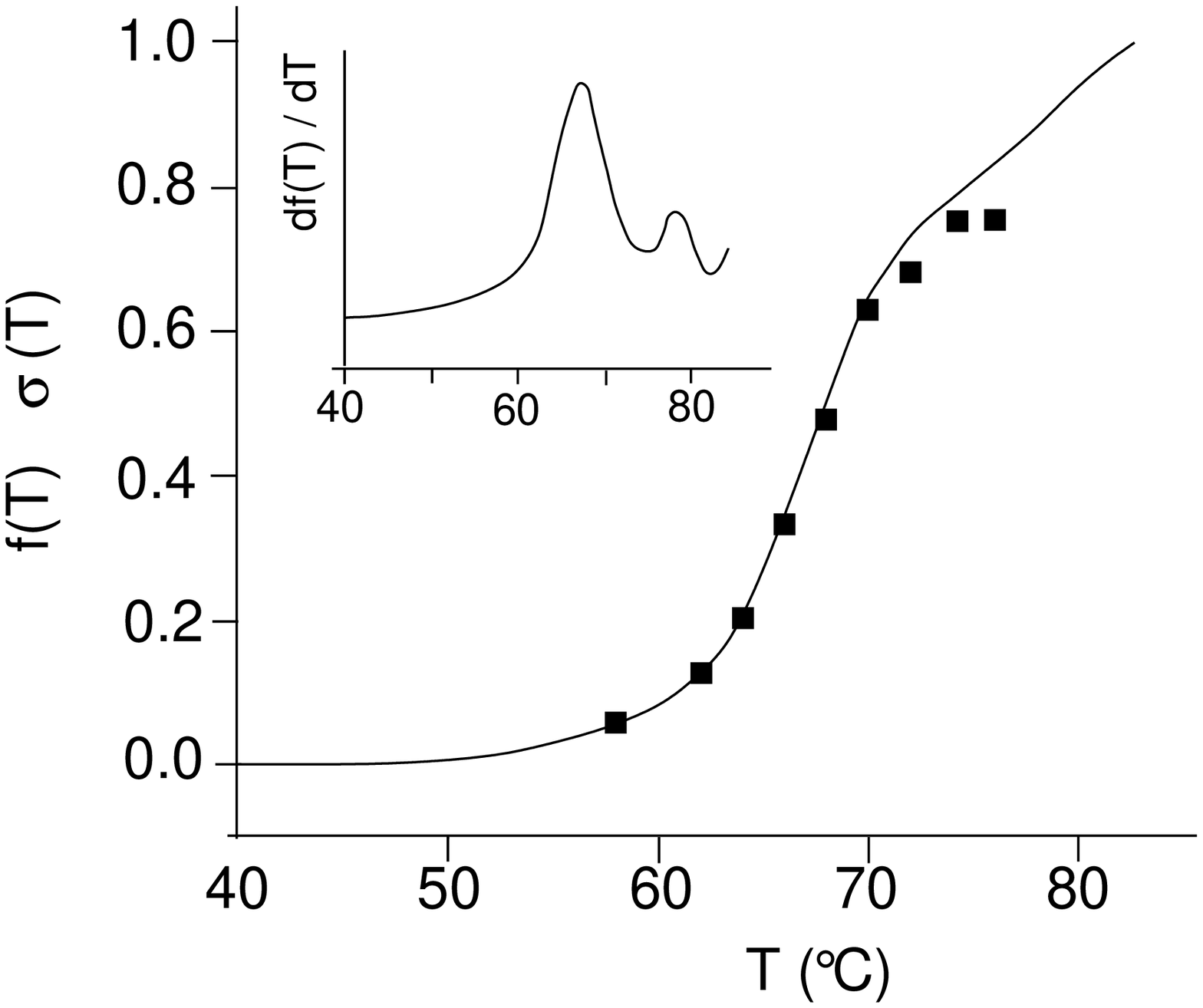} & 
\includegraphics[height=5cm]{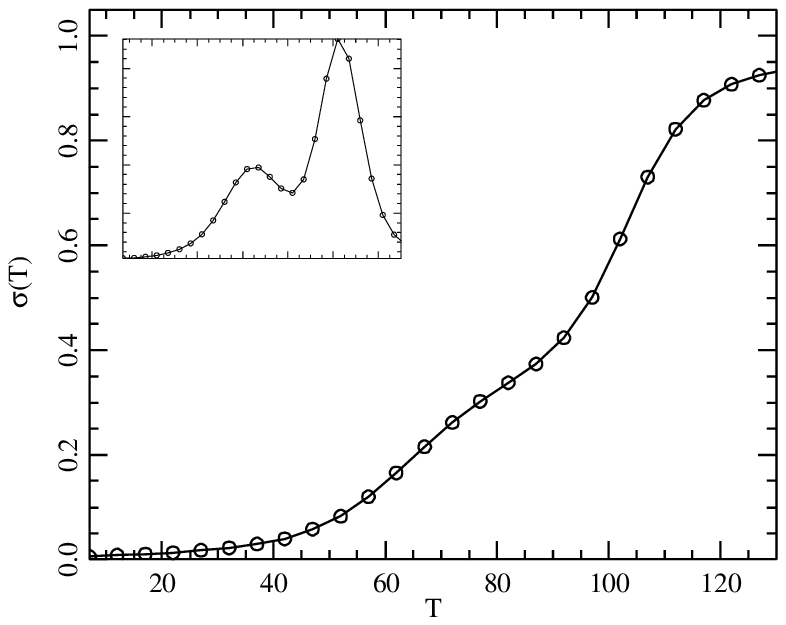} \\     
    \end{tabular}
  \end{center}
  \caption{(a) The experimental results obtained by Montrichok et
    al. \cite{Montrichok,ZENG} for the sequence L48AS shown in
    Fig.~\ref{fig:seq48}. Redrawn from \cite{Montrichok}.
   The line indicates the melting curve as it
    is observed by UV spectroscopy ($f(T))$ and the points show the
    fraction of open pairs that belong to a molecule which is not
    fully melted (dsDNA melting curve ($\sigma(T)$). 
     The inset plots the derivative  $df/dT$,
     with the same temperature scale as in the main figure, 
     to show the two-stage melting more clearly.
    (b) The dsDNA melting curve calculated with our model for the
    parameters listed in Table~\ref{table:param}. The inset shows the
    derivative of this curve versus temperature. The first peak
    corresponds to the melting of the AT-rich region (base pairs 1 to
    21, i.e. 40\% of the sequence) 
   and the second peak is associated to the melting of
  the larger GC-rich domain.}
\label{fig:montrichok48}
\end{figure}

This sequence, which has 48 base pairs, is shown in
Fig.~\ref{fig:seq48}. With its complementary strand, the single strand
shown in Fig.~\ref{fig:seq48} makes the double helix that was studied
in the thermal denaturation experiment. However the region extending
from base 22 to base 48 is such that it can also fold as a hairpin
formed of a 13-base pair stem closed by a single Adenine base. This
allowed the authors to measure both the fraction $f(T)$ of open pairs
versus temperature but also the fraction $\sigma(T)$ of open pairs
that belong to a molecule which is not fully melted (dsDNA melting
curve). The results of Montrichok et al. \cite{Montrichok,ZENG} are
shown in Fig.~\ref{fig:montrichok48}-a. Besides the part of the
sequence which was designed to make the hairpin and is mostly made of
GC base pairs, the remaining of the sequence (base pairs 1 to 21) is
mostly constituted of AT pairs. Therefore it is natural to expect that
base pairs 1 to 21 (about 40\% of the pairs) 
will open first at low temperature, and that the second
part will open at higher temperature. Therefore the denaturation curve
that one predicts for this sequence if a two-stage opening with a
plateau for an open fraction around $0.4$. This is exactly what Ising
models find when one uses the standard programs \cite{DINAMELT},
and this is also what our model gives,  as shown in
Fig.~\ref{fig:montrichok48}, although the separation between
the two stages is less
noticeable than for the Ising models. The three-dimensional picture of the
melting of this sequence, obtained from the model shows this expected
behavior, the opening starting in the AT-rich part of the sequence
and then stopping when it reaches base-pair 21, i.e. the beginning of
the GC-rich domain.

\begin{figure}
  \begin{center}
    \includegraphics[width=9cm]{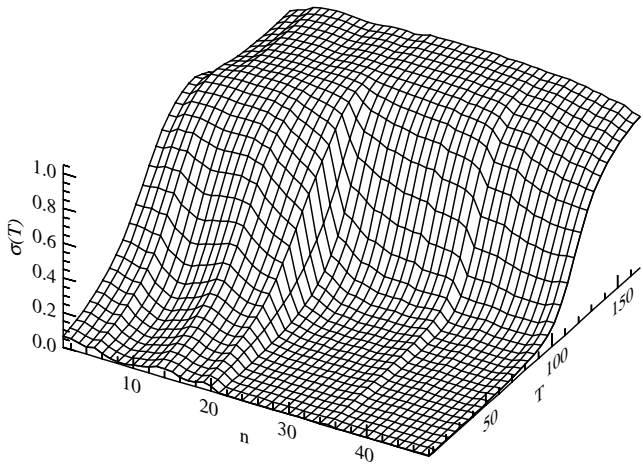}
  \end{center}
\caption{The three-dimensional dsDNA melting curve given by the model for
  the sequence shown in Fig.~\ref{fig:seq48}. The probability
  that a given base pair is open at a particular temperature is shown
  as a function of the index of the base pair. Indices are those
  listed in Fig.~\ref{fig:seq48}, the AT-rich region starting at index
1.}
\label{fig:3Dlocal}
\end{figure}

However the comparison with the
experimental results (Fig.~\ref{fig:montrichok48}
shows that all theoretical approaches are {\em
  qualitatively wrong!} It is important to stress that this is not a
simple weakness of a particular model because all Ising models,
irrespectively of the parameter set that they use provided it is
consistent with experiments for simple sequences, as
well as our theoretical model which is very different, face the same problem.
Moreover the discrepancy between theory and experiment is not a small
effect. The experiment finds that, although the opening occurs in two
stages, the first stage extends well inside the GC rich region (up to
$\sigma \approx 0.75$ i.e.\ base pair 36). Therefore, to understand
this property one must look for a deep explanation rather than
considering that it could be minor problem due, for instance, to
incorrect parameters. This particular sequence, which seems to
challenge all theoretical models is therefore particularly
interesting. 

A solution to this challenge can be provided by the experiments that
we presented in Sec.~\ref{sec:experiments}. They indicate that an AT-rich
region, which fluctuates widely, is able to influence the opening of
base pairs which are at least 8 base pair away in the sequence. In the
case of the experiment on sequence $L48AS$ this effect could even be
larger because the molecule opens at one end, so that the AT-rich
region can be expected to fluctuate even more than in our experiments,
where it is clamped at both ends by one or several GC pairs. This
suggests that, in the presence of a large AT-rich region {\em the
  model should include non-local effects to describe its influence on
  parts of the sequence which are not right next to it.} In the
experiments discussed in Sec.~\ref{sec:experiments} we noticed a
lowering of the local melting temperature in the vicinity of
an AT-rich region, which is particularly strong right next to it and
weaker further away. In the model this effect can be described
phenomenologically by a local rescaling of the parameters. We know
that the melting temperature decreases when the strength of the
stacking interaction decreases \cite{PB,TDPtransition}. Moreover the
large fluctuations in an AT-rich region can be expected to disrupt the
stacking in the vicinity. This is why we decided to model the effect
of an AT-rich region by introducing a softening of the stacking
interactions in the neighboring regions. We selected the following
rules
\begin{itemize}
\item Only AT-rich regions which exceed a given size of $k$
  consecutive AT pairs are taken into account.
\item Inside the  AT-rich domain, the parameters are not modified. 
\item for all stacking interactions which are less than $n_1$
  bonds from the AT-rich region, the coupling constant $K$ is
  multiplied by a factor $c_0 < 1$,
\item for all interactions which are $n$ bonds away from the
  AT-rich region, $n_1 > n > n_2$, $K$ is multiplied by $c_1 = c_0 + (1-c_0)(n -
  n1)/(n_2 - n_1)$ in order to reach smoothly the value 1 $n_2$ bonds
  away from the AT-rich region.
\item for all stacking interactions which are at least $n_2$ bonds
  away from the AT-rich region, the parameters of the interaction
  potential are not modified.
\item if a stacking bond is affected by more than one AT-rich region,
  which could be the case for the stacking interactions which are
  between two AT-rich domains, only the largest correction is
  considered.
\end{itemize}
From our experimental data on various sequences \cite{CuestaExp} and from
the theoretical studies of several cases, we set $k=9$, $n_1=10$, $n_2
= 16$, $c_0=0.2$, but these parameters could certainly be refined by
more extensive experimental studies.

\begin{figure}
\begin{center}
  \begin{tabular}{cc}
\textbf{(a)} & \textbf{(b)} \\
\\
\includegraphics[height=5cm]{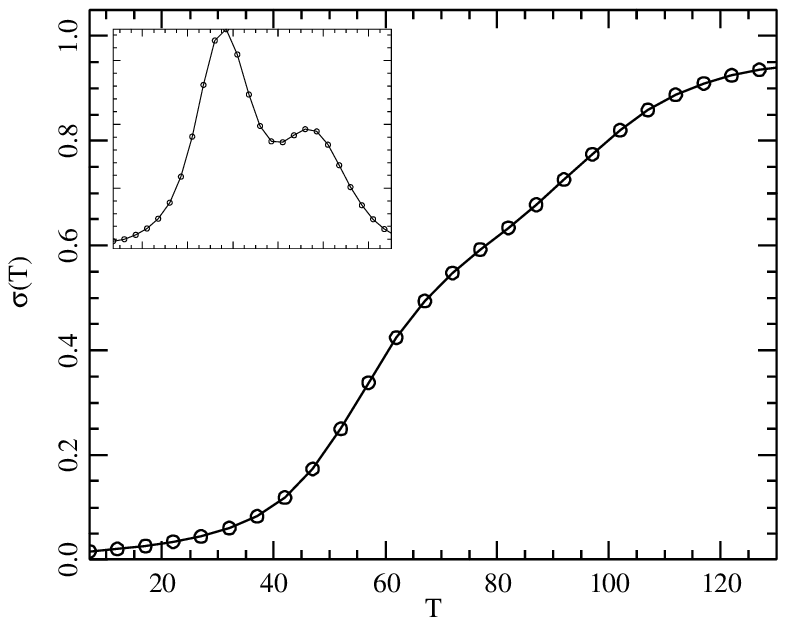} &
\includegraphics[height=5cm]{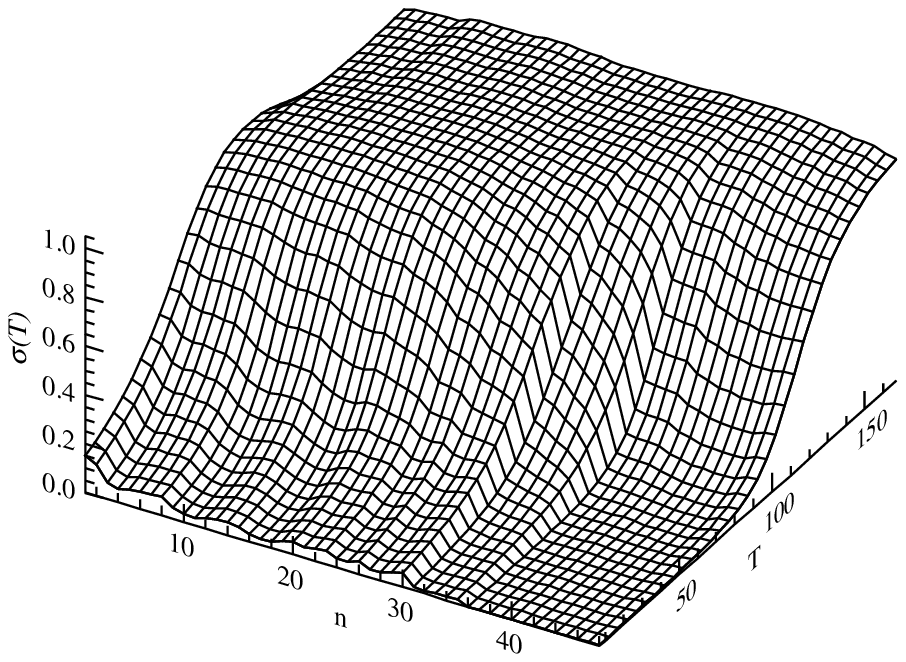}
  \end{tabular}
\end{center}
  \caption{Theoretical melting profile of the sequence $L48AS$ when
    the effect of the large AT-rich region on the stacking
    interactions along the sequence is taken into account. (a) dsDNA
    melting curve. The inset shows the derivative of the versus
    $T$, with the same temperature scale as in the main figure.
    (b) The corresponding three-dimensional dsDNA melting curve 
    given by the model. }
\label{fig:open48nonlocal}
\end{figure}

Figure \ref{fig:open48nonlocal} shows the theoretical melting curve
when the effect of the large AT-rich region is taken into account. The
derivative of the dsDNA melting curve still shows a two-stage melting
but, contrary to the first calculation (Fig.~\ref{fig:montrichok48}),
{\em the first stage is now the dominant one, in agreement with
experiments.} The three-dimensional melting curve indicates that, in
the first stage, the
melting that starts in the AT-rich end of the molecule extends inside
the GC region until approximately 75\% of the molecule, and the last
part opens then at higher temperature. The results are now in fairly
good quantitative agreement with experiments, although the transition
given by the model is still not as sharp as the experimental
curve. This could certainly be improved by optimizing the parameters,
but the one-dimensional model is affected by strong finite-size
effects which tend to broaden the denaturation transitions for short sequences.
Nevertheless the results exhibit a
qualitative change with respect to a calculation which only considered
local properties of the molecule to determine the melting
profile. This is an important point because this improvement cannot be
achieved by adjusting parameters while simultaneously preserving the
correct melting temperatures of AT or GC homopolymers. 

\begin{figure}
\begin{center}
  \begin{tabular}{cc}
\textbf{(a)} & \textbf{(b)} \\
\\
\includegraphics[height=5cm]{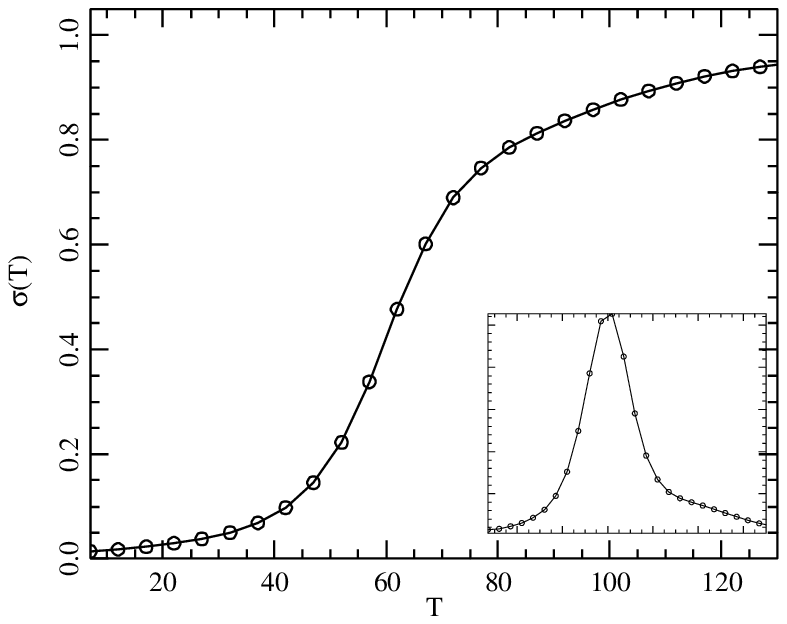} &
\includegraphics[height=5cm]{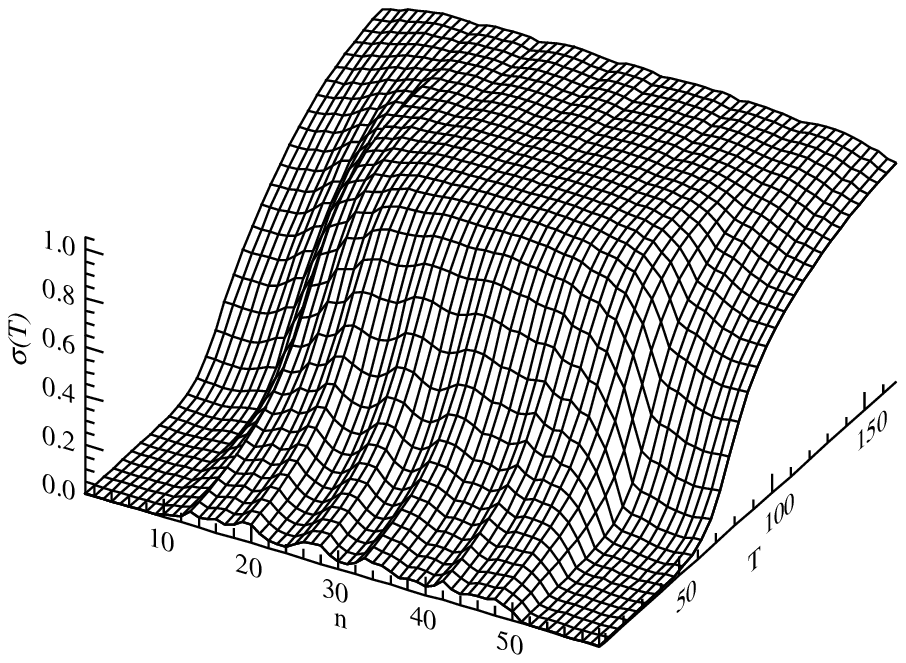}
  \end{tabular}
\end{center}
  \caption{Theoretical melting profile of the sequence $L60B36$
    \cite{ZENG}. The inset shows the derivative of the versus
    $T$, with the same temperature scale as in the main figure.
    (b) The corresponding three-dimensional dsDNA melting curve 
    given by the model. }
\label{fig:open60}
\end{figure}

\bigskip
The sequence $L48AS$ was particularly challenging. Most of the other
short sequences are much less sensitive to subtle non-local effects,
and therefore less demanding on the model to reach a satisfactory
agreement between theory and experiments. This is for instance the
case for sequence $L60B36$ of Ref.~\cite {ZENG} which has 60 base
pairs. It was designed with a 36 base pairs AT-rich region in the middle,
sandwiched by two GC rich domains. The AT-rich center is however
interrupted by a few GC pairs, the largest consecutive span of AT base
pairs containing only 8 pairs. According to our experiments
\cite{CuestaExp} a few isolated GC pairs scattered in an AT domain are
able to reduce significantly the fluctuations of this domain. This is
certainly the case for sequence $L60B36$, which explains why the small
GC rich domains at the ends of the sequence are sufficient to 
stabilize the molecule and prevent a full separation of the strands
in spite of the large open domain. Figure  \ref{fig:open60}
shows the melting profile given by the model, with the same parameters
as for the sequence $L48AS$. The three-dimensional profile clearly
shows the ``bubble in the middle'' that was detected in the
experiments \cite{ZENG}.

\section{Discussion}

The structure of the melting curve of DNA molecules of a few tens of
base pairs contains information at a scale which is relevant for the
genetic code, and this why its analysis is very interesting.

In this work we showed that it is possible to add a spatial dimension
to the melting curve through experiments that record the state of the
guanines along the sequence. This is still incomplete because only one
type of bases can be monitored but it nevertheless brings useful data
on the interplay between different parts of the molecule during the
melting. In particular the results indicate that an AT-rich region,
which fluctuates widely even at moderate temperature, can affect the
fluctuations of the double helix several base pairs away.

This experimental result could be questioned because it is
based on a  measurement
of the local fluctuations of the guanines which is not a direct
observation, but involves complex transformations of the DNA molecule,
including electron transfers and further chemical reactions. But it is
important to notice that a theoretical analysis of independent
measurements converges to the same conclusion.

For this theoretical analysis we significantly modified the nonlinear
model of DNA that was introduced earlier. Adding a barrier for closing
to the intra-pair potential $V$ is essential to get correct time
scales for the lifetime of the open states. As long as one computes
statistical quantities such as the mean stretching of base pairs at a
given temperature, it could seem irrelevant to care about
dynamics. But this is however an important information which should
not be neglected, and which is useful even for the calculation of
static equilibrium properties because the model with a modified
potential has different properties, such has a sharper melting
transition. The second evolution of the model was to introduce the
sequence in the stacking interactions because these ineractions depend
even more drastically from the sequence than the intra-base
potential.

The analysis of the melting curve of the sequnce $L48AS$, which
is challenging because standard Ising models, as well as our model
restricted to local effects, fail
to give a correct melting curve, even at a qualitative level,
showed the need of the introduction of the non-local effects 
of an AT-rich region to get satisfactory results. We described them
phenomelogically by a rescaling of the stacking interactions. This may
not be the only choice, and further studies combining theory and
experiments on various sequences are certainly necessary.

\ack Part of this work has been supported by R\'egion R\^one-Alpes
through its program CIBLE.

\end{document}